\documentclass[fleqn,twoside]{article}
\usepackage{espcrc2,amsmath,epsfig}

\newcommand{\prd}{Phys.\ Rev.\ D}
\newcommand{\rf}[4]{{#1} {\bf #2}, (#4) #3}

\newcommand{\physrl}{Phys.\ Rev.\ Lett.}

\newcommand{\np}{Nucl.\ Phys.}

\newcommand{\etal}{\emph{et al.}}

\newcommand{\mhat}{\widehat{m}} 
\newcommand{\psibar}{\overline{\psi}}

\newcommand{\Lqcd}{\ensuremath{\Lambda_{\text{QCD}}}}

\newcommand{\cond}{\langle\psibar\psi\rangle}

\newcommand{\oa}[1]{\ensuremath{{\cal O}(a^{#1})}}

\newcommand{\eref}[1]{Eq.~(\ref{#1})}

\title{Modelling the quark propagator}
 
\author{Patrick O.~Bowman\address[FSU]{Department of Physics and CSIT, 
Florida State University, Tallahassee FL 32306-4120, USA}\thanks{Talk presented
by POB.},
Urs M.~Heller\addressmark[FSU],
Derek B.~Leinweber\address[ADL]{CSSM and 
Department of Physics and Mathematical Physics,
University of Adelaide, Australia 5005} and
Anthony G.~Williams\addressmark[ADL]}

\begin{document}

\begin{abstract}
The quark propagator is at the core of lattice hadron spectrum calculations
as well as studies in other nonperturbative schemes.  We investigate the
quark propagator with an improved staggered action (Asqtad) and an improved
gluon action, which provides good quality data down to small quark masses.
This is used to construct ans\"{a}tze suitable for model hadron calculations
as well as adding to our intuitive understanding of QCD.
\end{abstract}

\maketitle

\section{The lattice quark propagator}

The quark propagator is a fundamental quantity of QCD.  Though gauge 
dependent, it manifestly displays dynamical chiral symmetry breaking, 
contains the chiral condensate and $\Lambda_{\text{QCD}}$, and has
been used to compute the running quark mass~\cite{Aok99}.  Some model 
hadron calculations rely on ans\"{a}tze for the quark propagator~\cite{Oettel},
yet on the lattice we have the opportunity to study it in a direct,
nonperturbative fashion.

We use the ``Asqtad'' quark action~\cite{Org99}, a highly improved staggered 
action that formally has no \oa{2} errors.  We extend some earlier 
work~\cite{staggered} by also using an improved gluon action.
We have calculated the quark propagator on three sets of configurations: 
$12^3 \times 24$ and $16^3 \times 32$ at $\beta = 4.60$ ($a = 0.125$ fm) and
$16^3 \times 32$ at $\beta = 4.38$ ($a = 0.167$ fm), each ensemble consisting
of 100 configurations. The configurations were fixed to Landau gauge.
Most results shown here are from the larger, finer lattice, where we used
8 quark masses: $ma$ = 0.012, 0.018, 0.024, 0.036, 0.048, 0.072, 0.108, 0.144
(19 to 114 MeV).  

In the (Euclidean) continuum, Lorentz invariance allows us to decompose the 
full quark propagator into Dirac vector and scalar pieces
\begin{equation}
S^{-1}(p^2) = Z^{-1}(p^2) [i \gamma \cdot p + M(p^2)].
\end{equation}
Asymptotic freedom means that, as $p^2 \rightarrow \infty$, 
$S^{-1}(p^2) \rightarrow  i\gamma \cdot p + m,$ (the free propagator)
where $m$ is the bare quark mass.

\begin{figure}[h]
\begin{center}
\epsfig{figure=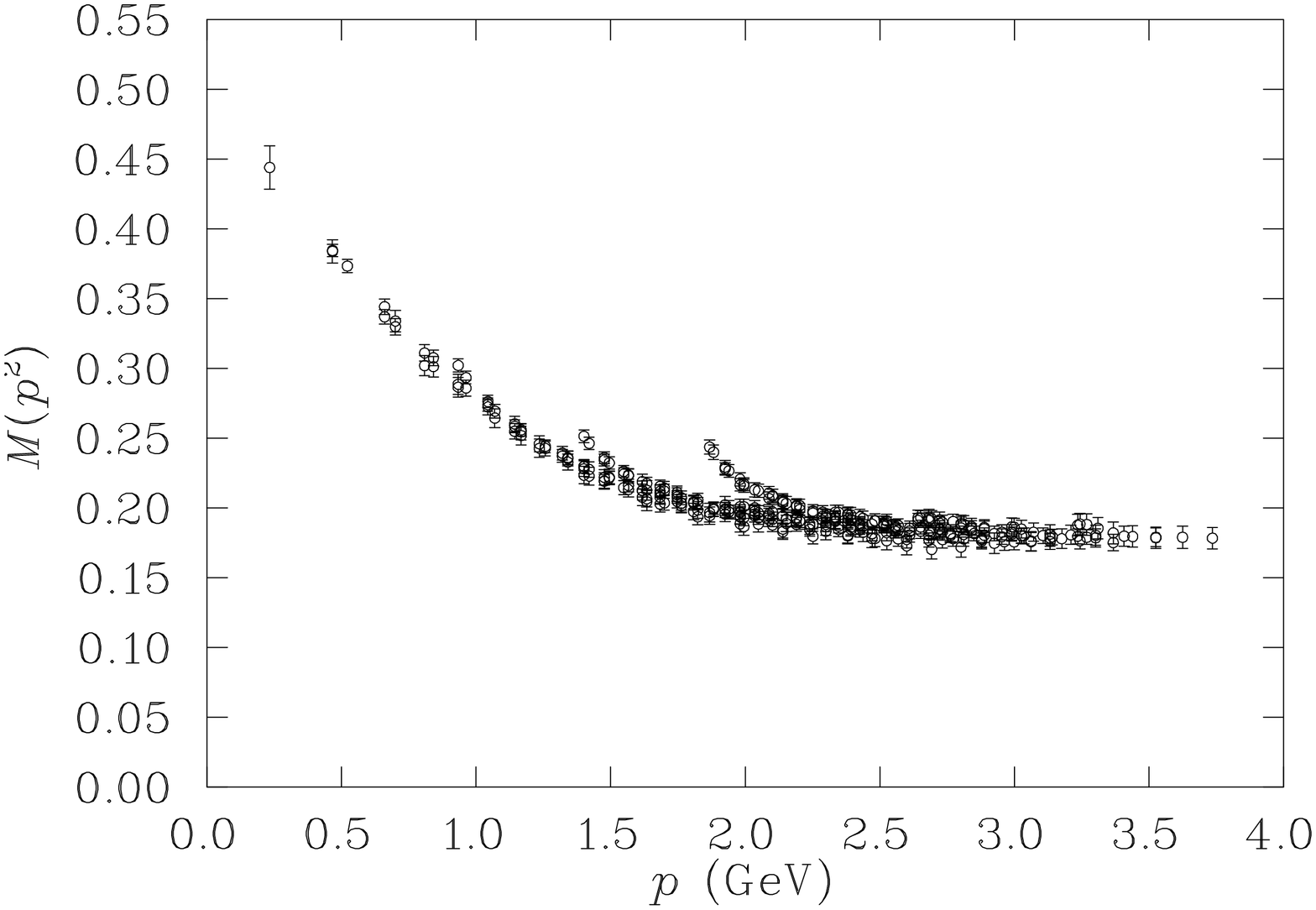,angle=90,width=7cm}
\end{center}
\begin{center}
\epsfig{figure=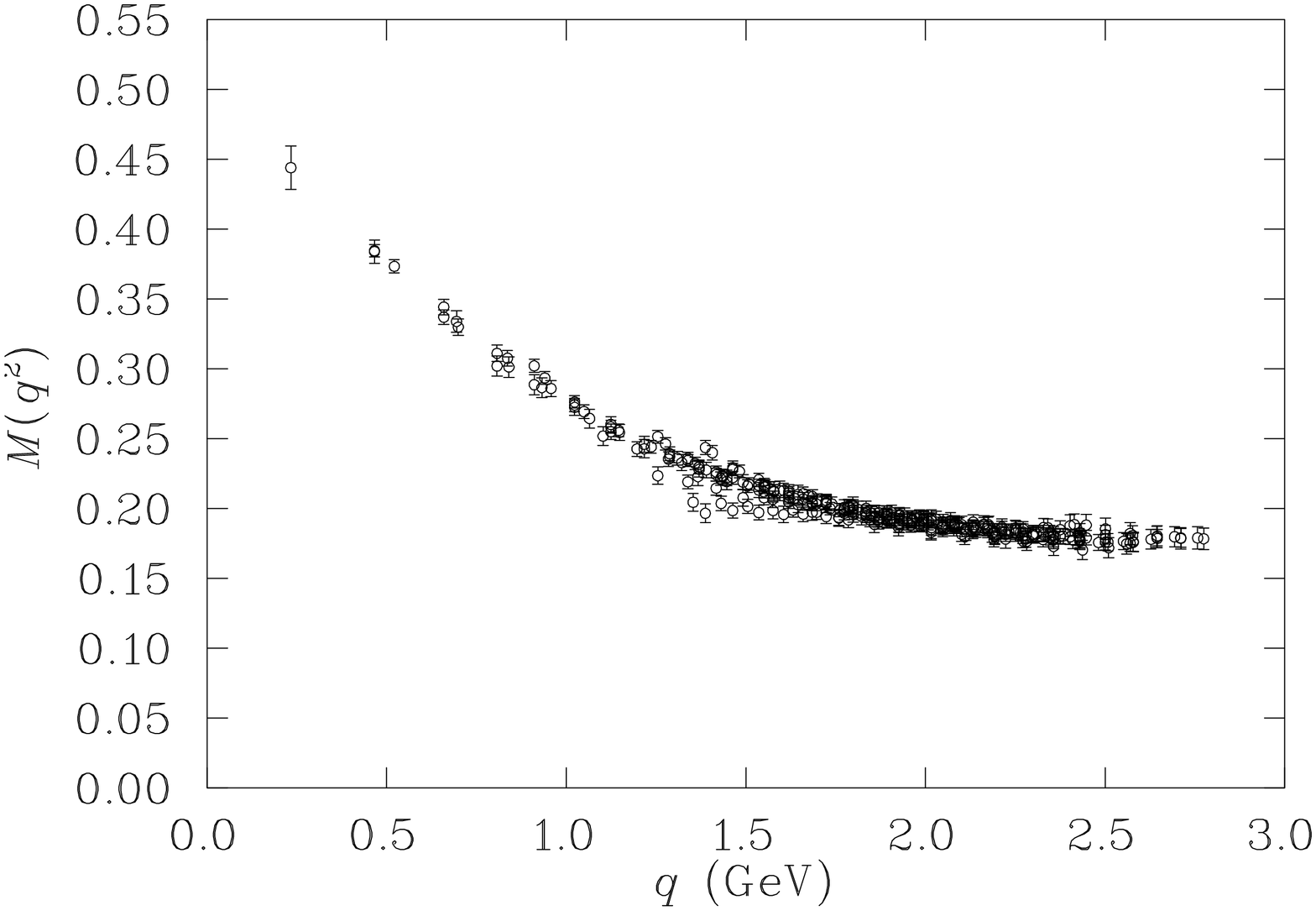,angle=90,width=7cm}
\end{center}
\vspace{-10mm}
\caption{Quark mass function at $\beta = 4.38$ ($a \simeq 0.167$ fm) for bare 
quark mass $m \simeq 114$ MeV. 
Top figure uses the na\"ive lattice momentum, while the bottom figure uses
the momentum derived from the tree-level behaviour of the action.}
\label{fig:quarkmom}
\vspace{-7mm}
\end{figure}

From consideration of the tree-level form of our lattice action, we
define the momentum variable 
\begin{equation}
q_\mu \equiv \sin(p_\mu) \bigl[ 1 + \frac{1}{6} \sin^2(p_\mu) \bigr]
\end{equation}
for the Asqtad action, where $p_\mu$ is the usual lattice momentum,
\begin{equation}
p_\mu = \frac{2\pi n_\mu}{aL_\mu} \qquad n_\mu \in 
	\Bigl[ \frac{-L_\mu}{4}, \frac{L_\mu}{4} \Bigr).
\end{equation}
By considering the propagator as a function of $q_\mu$ instead of $p_\mu$, we 
ensure that the lattice quark propagator has the correct tree-level form 
for $Z$ and hopefully better approximates its continuum behaviour.  It has 
already been seen that this produces a significant improvement in the 
rotational symmetry of $Z$~\cite{staggered}.  This prescription does not, 
however, tell us how to treat the mass function and in the past it has been
considered as a function of either $p$ or $q$ in different studies.  

Figure~\ref{fig:quarkmom} 
shows the quark mass function for the coarser lattice as a function of both $p$
and of $q$.  $M$ clearly has less anisotropy when plotted against $q$.  This
issue has been studied for the overlap quark propagator and there the
mass function appears to converge to the continuum limit most rapidly as a
function of $p$~\cite{Jianbo}.  The optimal momentum variable must therefore
be determined for each action.

\begin{figure}[h]
\begin{center}
\epsfig{figure=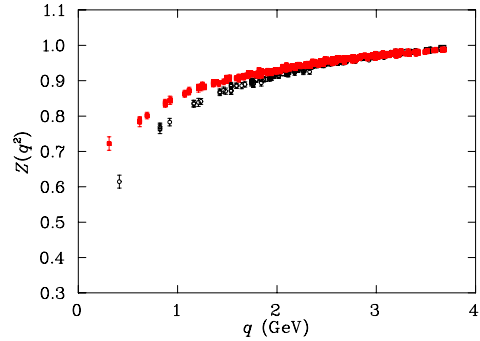,angle=90,width=7cm}
\end{center}
\vspace{-10mm}\caption{Quark $Z$ function for mass $ma = 0.036$, at 
$\beta = 4.60$.  Comparison on $12^3\times24$ lattice (open circles) and
$16\times 32$ lattice (solid squares).  For $M(q)$ on the same lattices there
are no discernible finite volume artefacts.}
\label{fig:finite_volume}
\vspace{-7mm}
\end{figure}

We test for finite volume effects by comparing the quark propagator on the
two $\beta=4.60$ lattices in Fig.~\ref{fig:finite_volume}.
No sign of finite volume artefacts is seen in the mass function, but the
infrared supression of $Z$ is stronger on the smaller lattice.

\begin{figure}[h]
\begin{center}
\epsfig{figure=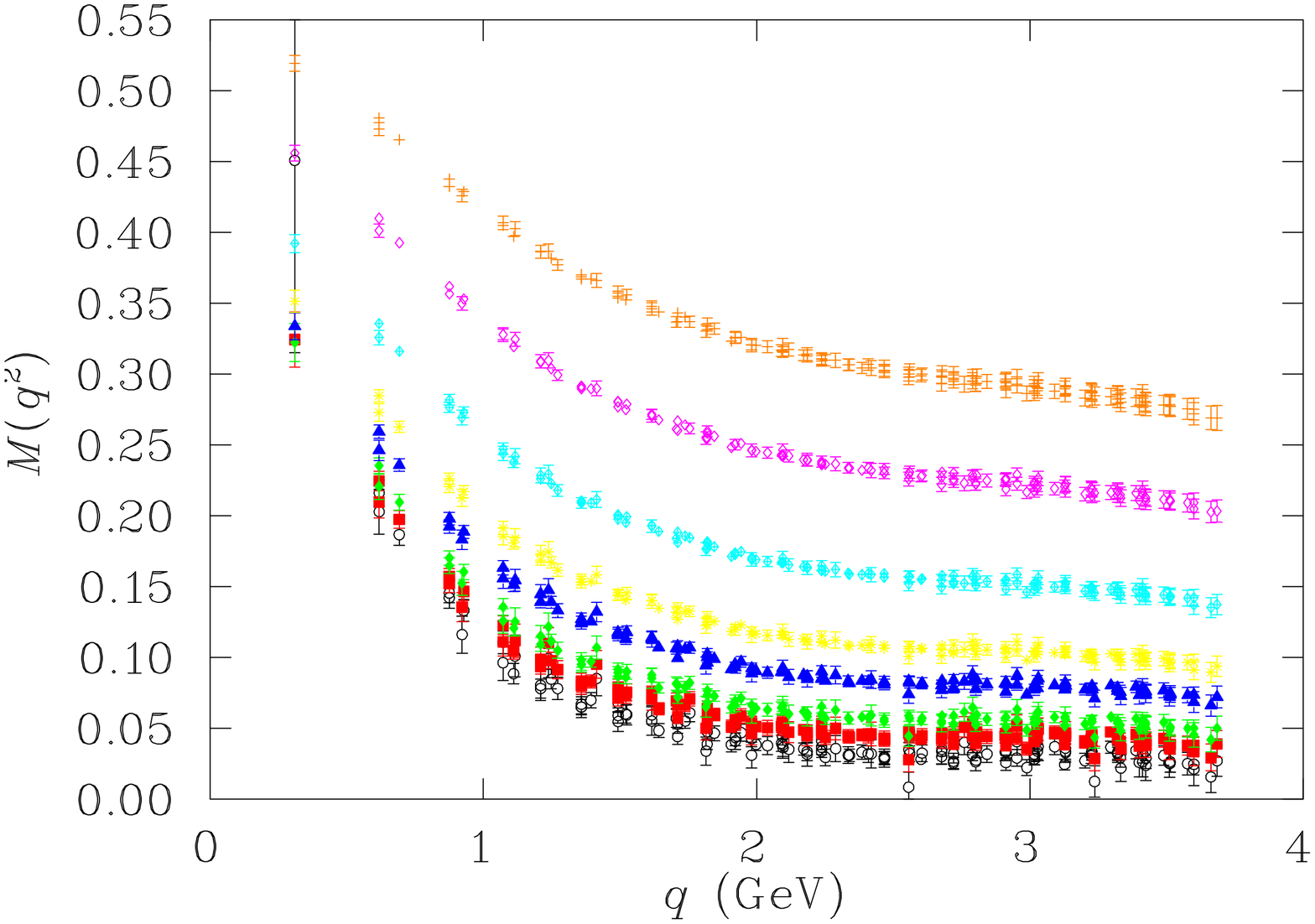,angle=90,width=8cm}
\end{center}
\begin{center}
\vspace{-5mm}
\epsfig{figure=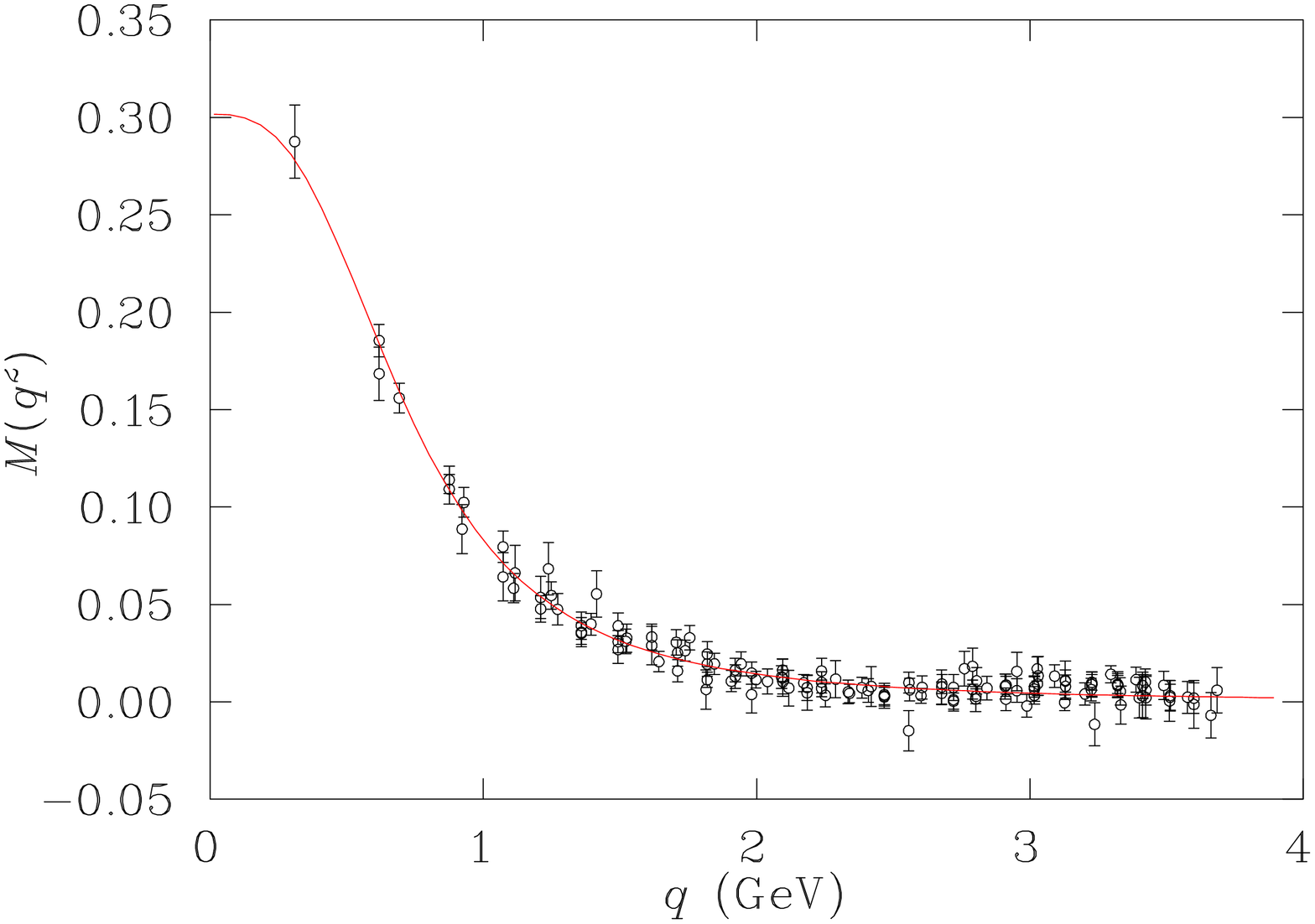,angle=90,width=7cm}
\end{center}
\vspace{-10mm}\caption{Quark mass function for all masses studied (top) and in
the chiral limit (bottom) on the larger $\beta = 4.60$ ensemble.  In the chiral
limit there is also a fit using the simple ansatz~\eref{eq:simplefit}.}
\label{fig:asq_b460_mall}
\vspace{-5mm}
\end{figure}

\section{Parameterising the data}

The mass dependence of $M(q)$ can be seen in Fig.~\ref{fig:asq_b460_mall}.
Dynamical chiral symmetry breaking can be seen in the infrared enhancement
of the mass function which, for small quark masses, is almost independent
of the bare mass.  For large quark mass, the function flattens off.
In the case of the $Z$ function, dependence on the quark mass is very weak.

\addtocounter{footnote}{-1}
In the continuum, in the chiral limit, the asymptotic behaviour of the quark 
mass function is related to the chiral condensate~\cite{Bon02}.
We performed an extrapolation to the chiral limit using a quadratic fit 
in the quark mass to the cylinder cut data.  A determination of the chiral
condensate was then be made by fitting to the tail of $M(q)$.  
We chose 2-3 GeV ($ap = 1.27 - 1.9$) for the fit region and set $\Lqcd$ at 
691 MeV\footnote{This was computed by taking the ALPHA result of 
$\Lambda^{\overline{MS}} = 239$ MeV~\cite{Cap99} and converting it to the
$\widetilde{MOM}$ scheme as described in Ref.~\cite{Bec99}.}.
We find $(-\cond)^{1/3} = 270(27)$ MeV; the quoted error is purely statistical.

 
A fit to each of the mass functions was done using the ansatz 
\begin{equation}
\label{eq:simplefit}
M(q^2) = \frac{c\lambda^{2(\alpha-1)}}{q^{2\alpha} + \lambda^{2\alpha}} 
         + \mhat.
\end{equation}
This is the same as the ansatz used in Ref.~\cite{staggered}, but written 
slightly differently. We shall refer to it here as the ``simple'' ansatz.  
Fit results are shown in Table~\ref{tab:simple_fits}.

\begin{table*}[t]
\caption{Best-fit parameters for the simple ansatz, 
\eref{eq:simplefit}. 
Statistical uncertainties are determined by jackknife.}
\label{tab:simple_fits}
\vspace{2mm}
\begin{tabular}{cccccccc}
\hline
$am$  & $a^3c$ & $\lambda$ (MeV) & $\mhat$ (MeV) & $\alpha$ 
& $M(0)$ (MeV) & $\chi^2$ / d.o.f. \\
\hline
 0     &  0.040(3) &  719(50) &   0    & 1.47(11)& 302(33) & 0.72 \\
 0.012 &  0.043(2) &  777(35) &  25(2) & 1.51(2) & 308(19) & 0.70 \\
 0.018 &  0.044(2) &  795(28) &  37(2) & 1.50(2) & 311(13) & 0.65 \\
 0.024 &  0.045(2) &  820(29) &  49(2) & 1.48(2) & 313(11) & 0.60 \\
 0.036 &  0.046(2) &  802(27) &  70(3) & 1.31(1) & 353(10) & 0.53 \\
 0.048 &  0.050(3) &  838(27) &  92(3) & 1.28(1) & 370(9)  & 0.48 \\
 0.072 &  0.059(2) &  946(29) & 138(5) & 1.30(8) & 397(12) & 0.44 \\
 0.108 &  0.071(5) &  997(27) & 199(6) & 1.12(3) & 478(4)  & 0.37 \\
 0.144 &  0.090(9) & 1100(32) & 255(10)& 1.01(5) & 547(4)  & 0.35 \\ 
\hline
\end{tabular}
\normalsize
\end{table*}

As before, we see that at small quark masses $\alpha \simeq 1.5$ is favoured,
while it is near one for the heavier quarks.  In this model, $\alpha$ is acting
 as a function of the bare mass, controlling the dynamical symmetry breaking.  
One fit is shown, for $M(q^2)$ at zero quark mass, in 
Fig.~\ref{fig:asq_b460_mall}.

We also investigated a more ``complete'' ansatz,
\begin{multline}
\label{eq:complete_fit}
\hspace{-8mm} M(q^2) = c \biggl( \frac{A\lambda^{2(\alpha-1)}}
         {q^{2\alpha} + \lambda^{2\alpha}} + \frac{1}{q^2+\lambda^2}
	 \Bigl[ \ln{\frac{q^2+\lambda^2}{\Lqcd^2}} \Bigr]^{\gamma_M} \biggr) 
	 \\ + \frac{\mhat}{\bigl[ \ln \bigl( (q^2 + \lambda^2)/\Lqcd^2 \bigr) 
	     \bigr]^{d_M}},
\end{multline}
which is an extension of the one used above that has the
correct asymptotic behaviour.  $A$ is dimensionless, $\mhat$ is the RGI quark
mass, $d_M$ is the anomalous dimension of the mass, $\gamma_M = d_M-1$, and 
$c$ can be related to the scalar condensate~\cite{Bon02}. 
Unfortunately, the current data is insufficient to simultaneously determine 
$A$ and $c$.  It is possible to determine $c$ separately from the asymptotic
behaviour, as above, but the end product fits the data no better than the 
simple ansatz.

The work of UMH and POB was supported in part by DOE contract 
DE-FG02-97ER41022.  DBL and AGW acknowledge financial support from the
Australian Research Council.


\end{document}